# High-harmonic generation from metasurfaces empowered by bound states in the continuum

George Zograf[1,2 #], Kirill Koshelev[1,2#], Anastasia Zalogina[1#], Viacheslav Korolev[3#], Duk-Yong Choi[4], Michael Zürch[3,5], Christian Spielmann[3], Barry Luther-Davies[4], Daniil Kartashov[3], Sergey Makarov[2], Sergey Kruk[1], and Yuri Kivshar[1,2]

[1]Nonlinear Physics Center, Australian National University, Canberra ACT 2601, Australia
[2]Department of Physics and Engineering, ITMO University, St. Petersburg 197101, Russia
[3]Institute of Optics and Quantum Electronics, Friedrich-Schiller University Jena, 07743 Jena, Germany
[4]Laser Physics Centre, Australian National University, Canberra ACT 2601, Australia
[5]University of California, Berkeley, CA 94720-1460, USA

**The concept of optical bound states in the continuum (BICs) underpins the existence of strongly localized waves embedded into the radiation spectrum [1] that can enhance the electromagnetic fields in subwavelength photonic structures [2, 3, 4]. Early studies of optical BICs in waveguides [5, 6, 7] and photonic crystals [8] uncovered their topological properties [9, 10, 11], and the concept of quasi-BIC metasurfaces [12] facilitated applications of strong light-matter interactions to biosensing [13, 14], lasing [3, 15], and low-order nonlinear processes [4]. Here we employ BIC-empowered dielectric metasurfaces to generate efficiently high optical harmonics up to the 11$^{th}$ order. We optimize a BIC mode for the first few harmonics and observe a transition between perturbative and nonperturbative nonlinear regimes. We also suggest a general strategy for designing subwavelength structures with strong resonances and nonperturbative nonlinearities. Our work bridges the fields of perturbative and nonperturbative nonlinear optics on the subwavelength scale.**

[#] These authors contributed equally to this work

High-harmonic generation (HHG) was first observed in the decades immediately following the development of the first lasers. The generation of optical harmonics up to the 11[th] order was achieved in a plasma in 1977 [16]. Since then HHG has found applications as sources of extreme UV and soft X-rays [17, 18, 19]; as sources of attosecond pulses [20]; as well as for all-optical atomic-scale measurements. For a long time the field has been dominated by gas-based systems [21]. Gas-phase HHG, however, involves expensive vacuum set-ups and complicated methods to confine the source gas only within the interaction volume, thus limiting its applicability. More recently, HHG from solids has been demonstrated [22, 23] revealing rich new physics [24] and offering systems with more compact form-factors. The use of solid-state platforms allowed HHG to enter the realm of nanoscience, and several pathways to nanoscale HHG have been explored, including plasmonic nanostructures [25], as well as dielectric metasurfaces [26] and gratings [27].

Research on HHG on the subwavelength scale is linked to previous extensive developments of subwavelength resonators for lower-order nonlinearities [4, 28, 29]. However, there exists a fundamental difference between lower- and higher-order nonlinear processes. Lower-order nonlinear light-matter interactions can often be considered as small perturbations to the linear regime of interactions and their physical origin is the nonlinear polarization current in the ground state of the quantum system. In contrast, higher order nonlinearities often are fundamentally dependent on processes that go beyond the perturbative regime. Although details of the solid-state HHG mechanism remains a topic for debate [24], the process itself is related to strong field population transfer from the valence to the conduction band, and, therefore, involves the formation of an electron-hole plasma that can significantly alter the optical properties of the material during the interaction. As a result of high carrier density in the conduction band, subwavelength resonators designed for perturbative nonlinear interactions are expected to become sub-optimal in nonperturbative regimes, suggesting it is necessary to develop new design principles for nonperturbative nonlinearities that would take into account the dynamic changes in the index of refraction and absorption occurring on ultrafast time scales [30].

Here we bridge the emerging field of nonperturbative HHG with the well-understood perturbative nonlinear nanophotonics by employing designs based on bound states in the continuum (BIC). To this end, we design a set of BIC metasurfaces with variable radiative losses dependent on a single geometric parameter (asymmetry). We employ the concept of critical coupling determined by the balance of radiative and nonradiative losses of the BIC metasurfaces [31, 32]. Based on this approach, we report strong enhancement of odd optical harmonics (from 3[rd] to 11[th]) from a silicon metasurface hosting BIC, as illustrated in the conceptual image of Fig.1a.

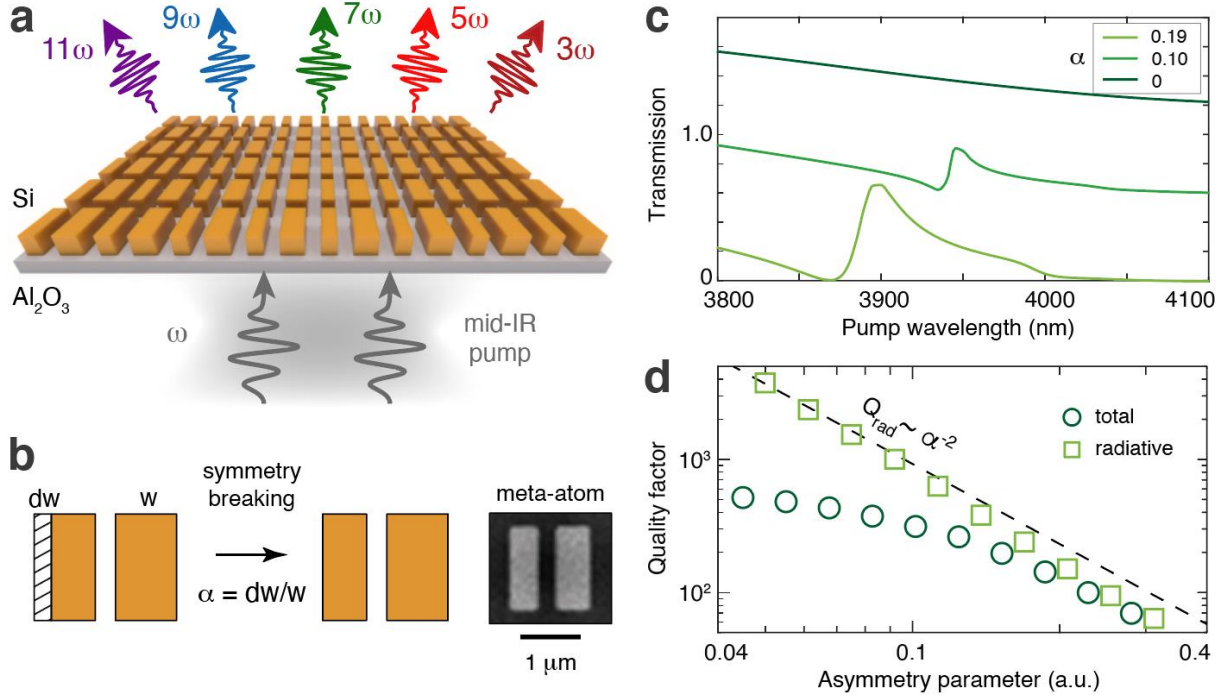

**Fig. 1. Metasurface supporting quasi-BIC resonances.** (a) Concept of a silicon metasurface generating 3$^{rd}$ to 11$^{th}$ optical harmonics. (b) Schematic and SEM image of the unit cell with broken in-plane symmetry. Definition of the asymmetry parameter $\alpha$. (c) Simulated transmission spectra for different values of $\alpha$. For clarity the spectra have been shifted relatively by 0.6 units along the vertical axis. (d) Characteristic dependence of the total (squares) and radiative (circles) Q factor of the quasi-BIC on the meta-atom asymmetry parameter α. For zero asymmetry parameter the total quality factor saturates at the value of the nonradiative quality factor.

In our study, we consider resonant dielectric metasurfaces composed of complex meta-atoms in the form of an asymmetric pair of rectangular silicon bars on a sapphire substrate, as shown in Fig. 1b. The bars in each pair have the widths $w$ and $w - dw$, respectively, and the asymmetry parameter is defined as $\alpha = dw/w$. We perform numerical analysis of the linear response and eigenmode spectra of the resulting asymmetric metasurface using commercially available software based on full-wave electromagnetic simulations (see details in Methods). We design a square lattice metasurface with the period of 2040 nm, thickness of 1150 nm, bar length of 1445 nm, larger bar width of 595 nm, and distance between bars of 510 nm. The width of the smaller bar is varied in the range 595 nm to 480 nm, which corresponds to the asymmetry parameter of $\alpha = 0$-$0.19$, respectively.

The calculated transmission spectra are shown in Fig. 1c. A symmetric metasurface with $\alpha = 0$ supports a BIC at 3978 nm with an infinite quality factor (Q factor) protected by the in-plane symmetry of the unit cell [1]. The presence of the BIC does not appear in the transmission spectrum at $\alpha = 0$ because its symmetry is incompatible with modes of free space. For the broken-symmetry metasurfaces with $\alpha = 0.19$ and $\alpha = 0.1$, the transmission curves reveal a sharp resonance with a Fano line shape associated with a quasi-BIC with a high Q-factor [12]. The radiative Q-factor of quasi-BICs in asymmetric metasurfaces follows the typical inverse quadratic dependence on the meta-atom asymmetry parameter, as shown in Fig. 1d. The total Q-factor of the quasi-BIC mode is limited by nonradiative losses such as material absorption, parasitic scattering due to surface roughness, lattice disorder and the finite extent of the sample [31].

We fabricate the BIC metasurfaces from an amorphous Si film on a sapphire wafer using electron beam lithography (see details in Methods). A scanning electron microscope (SEM) image of the resulting meta-atom is shown in Fig. 1b, while the metasurfaces images of larger scale are shown in Fig. 2a.

In our optical experiments, we pump the BIC metasurfaces with beams from various pulsed mid-IR optical parametric generators and detected signals at the harmonics of the pump using cooled near-IR and visible detectors (see details in Methods). We perform experiments with three different pump pulse durations: $\tau = 2$ ps, 160 fs and 100 fs. We observe the generation of optical harmonics up to the 5$^{th}$ order using the 2 ps pulses; up to the 7$^{th}$ order for 160 fs pulses; and up to the 11$^{th}$ order for 100 fs pulses. Figures 2e,f,g show the detected spectra for the 3$^{rd}$ to 11$^{th}$ optical harmonics for the optimal pump wavelength of 3805 nm. The harmonics show a strong dependence on the pump polarization as depicted in Fig. 2b (see also Supplementary Section S1). Figure 2d shows the scaling of the harmonic intensities as a function of pump intensity. We observe deviations of power scaling predictions from the perturbative model for all harmonics above the 3$^{rd}$. Namely, while the dependencies for the 3$^{rd}$ and 5$^{th}$ harmonics follow the laws $I_{3w} \sim I_w^3$ and $I_{5w} \sim I_w^5$, respectively, using 2-ps laser irradiation, we find that 5$^{th}$, 7$^{th}$, 9$^{th}$, and 11$^{th}$ harmonics may be best described by the dependences $I_{5w} \sim I_w^4$, $I_{7w} \sim I_w^{4.8}$, $I_{9w} \sim I_w^{5.2}$, and $I_{11w} \sim I_w^{3.8}$, respectively, when being generated by intense 100-fs pulses. Moreover, these dependencies exhibit saturation at the higher intensities pointing to a nonperturbative character of the HHG mechanism for intense ultrashort laser pulses.

Since the 5$^{th}$ harmonic can be observed for both 2-ps and 100-fs pump pulses (Fig. 2), we study the dependence of the enhancement factor for an optimal metasurface on the pulse duration assuming their spectrally limited character and keeping their intensity in the perturbative range (i.e. $10^{-4} - 10^{-3}$ TW/cm$^2$). Correlating with our modeling of Figs. 1c,d and previous studies of perturbative harmonics enhancement from quasi-BIC states [4, 31], the enhancement factor for 5$^{th}$ harmonics is found to be about 45 (Fig. 3a, blue curve). In turn, Fig. 3a reveals a considerable decrease of the maximum enhancement factor with shortening the pulse duration and, thus, its spectral broadening (11 times for 160 fs and 4 times for 100 fs).

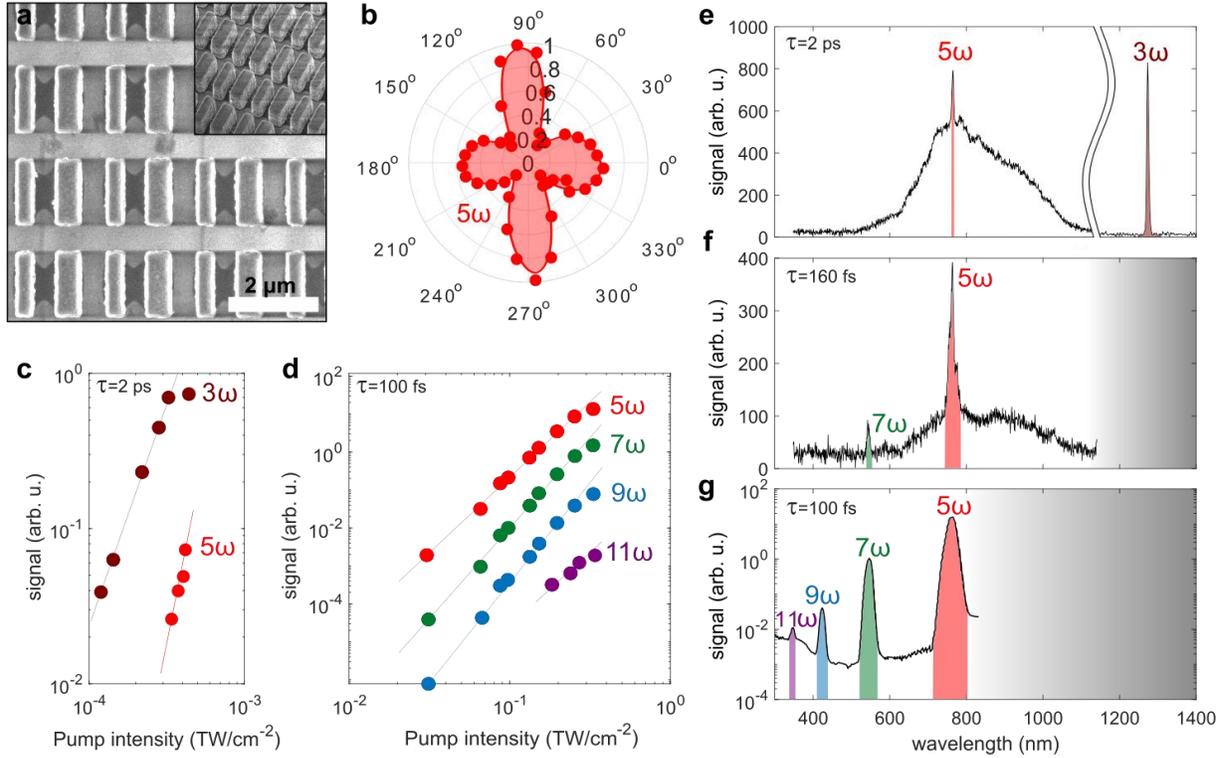

**Fig. 2. High-harmonic generation from an optimized BIC metasurface.** (a) SEM images of Si metasurface: top and side views. The scale bar is 2 μm. (b) Dependence of 5$^{th}$ harmonic intensity on the polarization angle of the linearly polarized pump with orientation of the metasurface corresponding to the top-view SEM image in (a). (c) Measured signals of 3$^{rd}$ and 5$^{th}$ harmonics versus the intensity for 2 ps pulse duration pump. (d) Measured signals of HHG versus the intensity for 100 fs pulse duration pump. (e) Spectrum of the third and fifth optical harmonics generated by laser pulses with duration of 2 ps. (f) Spectrum of the 5$^{th}$ and 7$^{th}$ optical harmonics generated by laser pulses with duration of 160 fs. (g) Spectrum of HHG generated by laser pulses with duration of 100 fs. For (f) and (g) the grey area corresponds to the spectral range that is not covered for a particular spectrometer. For (e) broken x-axis corresponds to different spectrometers used to detect 3$^{rd}$ and 5$^{th}$ optical harmonics.

Complementary to this experiment, we study numerically the enhancement factor on the asymmetry parameter of the BIC metasurfaces, and observe a dramatic reduction of the HHG efficiency for shorter pulses and higher intensities as shown in Fig. 3b. With these results, one can reach a paradoxical conclusion that shorter pulses and higher peak intensities may reduce the efficiency of the harmonics generation in subwavelength resonators. For example, the enhancement factor for 5$^{th}$ harmonics drops by more than one order of magnitude at the most resonant condition (quasi-BIC state) after the pulses compression by 20 times (Fig. 3a). In order to resolve this paradox, we recall the mechanism of HHG in the nonperturbative regime.

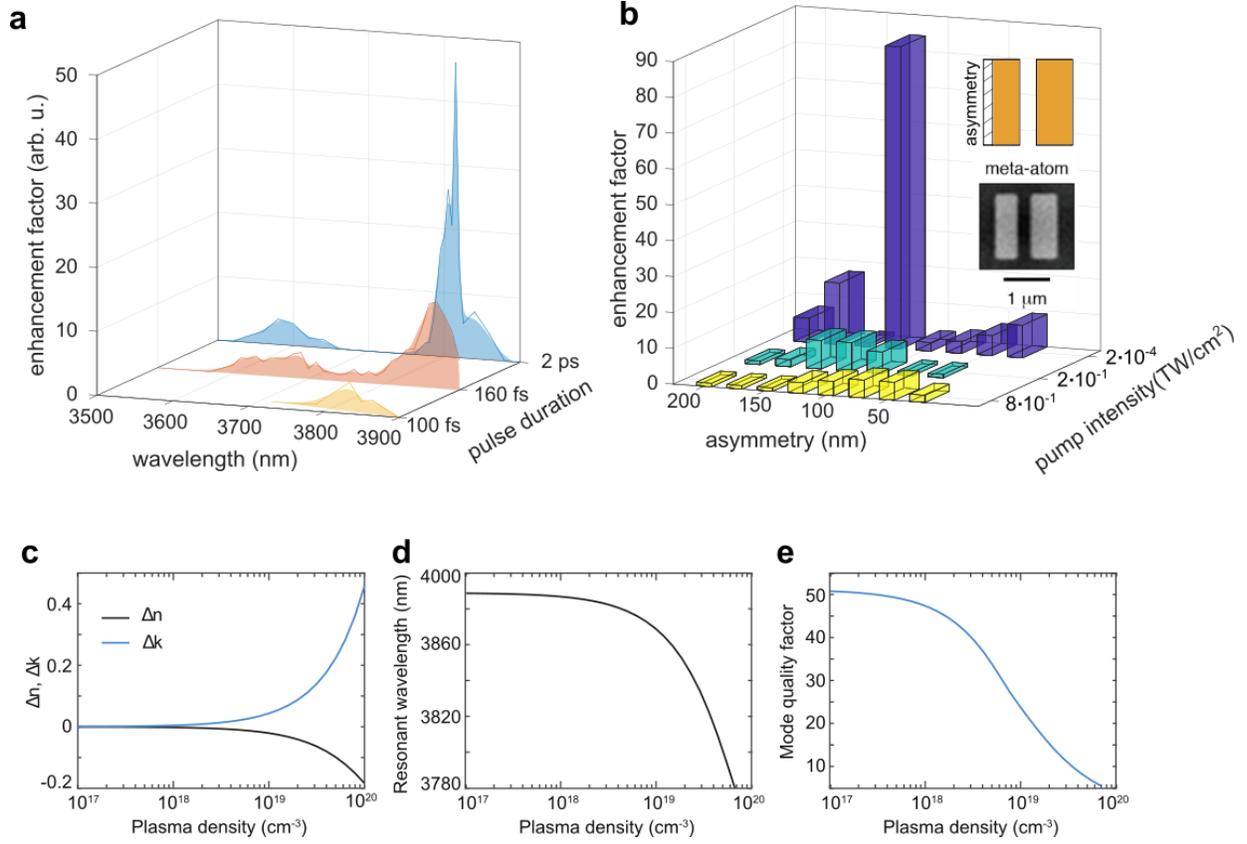

**Fig. 3. Transition from the perturbative to nonperturbative regimes**. (a) Experimentally measured dependence of 5$^{th}$ harmonic generation versus pump pulse duration. (b) Experimentally measured signal of harmonics generation versus asymmetry parameter at different pump pulse intensities. For intensities 0.8 and 0.2 TW/cm$^2$ the data is for 5$^{th}$ harmonic at $\tau$ = 100 fs. For intensity 2×10$^{-4}$ TW/cm$^2$ the data is for 3$^{rd}$ harmonic at $\tau$ = 2 ps. Calculated dependences of the material real $\Delta n$ and imaginary $\Delta k$ parts of refractive index (c), spectral position of the quasi-BIC resonant wavelength (d), and quasi-BIC Q-factor value (e) on the density of electron-hole plasma.

In the nonperturbative regime, the high harmonics originate from either intraband nonlinear current in the conduction band (nonlinear Bloch oscillations), or inter-band electron-hole recombination similar to the mechanism previously well established in gas-phase HHG [24]. According to the work [33], the intraband mechanism dominates over interband for harmonics with photon energies less than the band gap (e.g. 3.4 eV direct band gap in Si, i.e. up to 9$^{th}$ harmonics for 3.6-3.9 μm laser wavelengths used in our experiments). Thus, in our system, the transition from perturbative to nonperturbative regimes of nonlinear light-matter interactions is associated with generation of free carriers. Also, in an amorphous silicon film, one can still observe influence of indirect bandgap edge in the absorption spectrum around 1.1 eV (see further details in Supplementary, Section S6) and broadband photoluminescent background in Fig. 2g, as well as the absence of harmonics below the wavelength of 380 nm (3.3 eV), that suggests that the direct transition 3.4 eV is involved for coherent (phonon-free) HHG process.

The rate of free carrier generation is determined by the transition of electrons to the conduction band of Si via tunneling or multi-photon absorption processes. It can be evaluated by using the quantitative model, originally proposed in [34] and later generalized in [35]. An important parameter for describing the strong field ionization is the so-called Keldysh parameter

$$\gamma(t) = \omega (m/\Delta)^{1/2}/e |E(t)|, \quad (1)$$

where $\Delta = 3.4$ eV is the direct band gap of Si, $m = 0.18 m_e$ is the reduced mass of the effective electron and hole masses, $\omega$ is the pump frequency, $e$ is the elementary charge and $|E(t)|$ is the amplitude of the electric field inside the metasurface enhanced by the quasi-BIC resonance. Our calculations based on temporal coupled-mode equations for the field amplitude (for details, see Supplementary Sections S2-S3) show that at the hotspot of the electric field of the quasi-BIC the Keldysh parameter reaches its minimal value of $\gamma \sim 0.3$. Since this value is sufficiently less than the unity, strong field ionization is dominated by the tunneling process [35]. Numerical simulations using the explicit Keldysh model show that after excitation of the quasi-BIC resonance with a Gaussian pulse with the intensity of $3\times10^{11}$ W/cm$^2$ and duration of 100 fs, the electron-hole plasma density in the field hotspots reaches $n_p = 6.8\times10^{19}$ cm$^{-3}$ (for details, see Supplementary Section S4).

The electron-hole plasma induces a local change of the dielectric permittivity by the Drude correction with a plasma frequency of $\omega_p=1.1\times10^{15}$ rad/s. The plasma frequency becomes comparable with that of incident light ($\omega=0.483\times10^{15}$ rad/s). According to our estimation of the Drude term (for details, see Supplementary Section S5) for $n_p = 6.8\times10^{19}$ cm$^{-3}$ and Drude relaxation time of 1 fs, we get $\Delta n = -0.13$ and $\Delta k = 0.3$ (see Fig. 3c). Such large changes of refractive index induce a shift of the quasi-BIC resonant wavelength and decrease of its Q-factor, as shown in Figs. 3d, 3e. The wavelength decreases from 3890 nm to 3780 nm, while the Q-factor drops. The resonant properties of quasi-BICs are strongly suppressed for high densities $n_p$ and the resonance is smeared out. As a result of our calculations, high power-dependent concentrations of free carriers make the optical parameters of Si variable in the nonperturbative regime, while they remain virtually constant in the perturbative regime. Thus, an enhancement of HHG via critical coupling occurs at different levels of radiative loss for both perturbative and nonperturbative regimes.

In conclusion, we have employed the concept of optical bound states in the continuum to generate high harmonics with metasurfaces. We have designed and fabricated metasurfaces from amorphous silicon supporting a quasi-BIC resonance in the mid-IR spectral range. Being resonantly excited, the metasurface generates 3$^{rd}$ to 11$^{th}$ odd optical harmonics extending through the near-IR to the visible spectral ranges. The observed harmonic generation is strongly enhanced for the optimal metasurface geometry (characterized by the asymmetry parameter), as well as the pump wavelength and polarization. We have probed the strong dependence of $5^{th}$ high harmonic generation on the temporal properties of the pump. We have observed a dramatic efficiency enhancement and spectral narrowing of the high-harmonic generation for varying the pump conditions. We have traced the transitions from perturbative to non-perturbative regimes in nonlinear light-matter interactions and have demonstrated associated changes of the optimal value of the asymmetry parameter. The concept of resonant metasurfaces with highly localized light achieved by employing the BIC concept provides a new avenue to control experimentally strong nonlinear optical response of metasurfaces.

## METHODS

**Numerical and analytical methods.** For numerical simulations of the linear response, we used the finite-element-method solver in COMSOL Multiphysics in the frequency domain. The near-field distributions were simulated using the eigenmode solver in COMSOL Multiphysics. All calculations were realized for a metasurface placed on a semi-infinite substrate surrounded by a perfectly matched layer mimicking an infinite region. The simulation area was the unit cell extended to an infinite metasurface by using the Bloch boundary conditions. The material properties, including absorption losses, were extracted from the ellipsometry data for Si and imported from tabulated data for a sapphire substrate. The incident field was a plane wave in the normal excitation geometry, polarized along the long side of the bars. The non-radiative losses due to surface roughness were introduced artificially via addition of a 40 nm absorptive layer to the surface of bars with $k = 0.02$, where $k$ is the extinction coefficient.

To evaluate the time-dependent field enhancement at the quasi-BIC, we employed the temporal coupled-mode theory based on an explicit expansion of Maxwell's equations into the basis on orthogonal resonant states, for more details see Supplementary Sections S2-S3. For calculations of the strong field ionization rate and the generated electron-hole plasma density we used the Keldysh model, discussed in Supplementary Section S4.

**Sample fabrication**. The fabrication process of the BIC metasurfaces began with the deposition of an amorphous Si film on a sapphire wafer using plasma-enhanced chemical vapor deposition (Temescal BJD-2000). The thickness was measured using ellipsometry (JA Woollam M-2000D). For the next step, the photoresist layer ZEP 520A (1:2) Anisole was deposited. A thin dissipative layer of gold was evaporated using a thermal evaporator to prevent charging during pattern writing. Next, the metasurface pattern was defined via electron-beam lithography (Raith-150 EBL system), the area dose of the light explosion was 140 µAs/cm². Then the pattern was transferred onto the material using a reactive ion etching (inductively-coupled plasma) ICP-Fluorine (Samco 400iP) with $CHF_3$, $SF_6$, Ar and $O_2$ gases. Electron microscope images of the resulting sample were obtained using an FEI Verios scanning electron microscope.

**Optical experiments.** The 3$^{rd}$ and 5$^{th}$ optical harmonics generation from metasurface with linearly polarized pump at 2 ps (Fig. 2b,e and 3a) and 160 fs (Fig. 3a) pulse duration and 5 MHz repetition rate were obtained using a mid-infrared laser system consisting of a laser (Ekspla Femtolux 3) 5$^{th}$, and an optical parametric amplifier (MIROPA from Hotlight Systems). To detect the signal, we used a pair of peltier-cooled spectrometers working in visible-to-near- infrared range (QE Pro by Ocean Optics) and in near-infrared range (NIR-Quest by Ocean Optics). High optical harmonics generation spectra (5$^{th}$ to 11$^{th}$) that are shown in Figs. 2c,d,f and Figs. 3a,b were obtained using linearly polarized mid-IR pulses of duration of 100-fs and a repetition rate of 1 kHz from a Ti:Sapphire laser (Coherent Astrella) pumped optical parametric amplifier (TOPAS). The signal detection was done with the Andor Kymera 328i spectrometer system equipped by UV-enhanced, cooled CCD camera as a detector.

**Acknowledgements**

The authors acknowledge a financial support from the Australian Research Council and the Russian Foundation for Basic Research (grants 19-32-90106, 19-02-00419), the Federal Ministry of Education and Research (BMBF) under the "Make our Planet Great Again – German Research Initiative" (grant 57427209) implemented by the German Academic Exchange Service (DAAD), and by the German Research Foundation DFG (CRC 1375 NOA). The authors thank Richard Hollinger for help with alignment of the optical setup.

**Author contribution**

S.K and Y.K. conceived the idea, S.M., D.K., S.K. and Y.K supervised the research of three groups. G.Z., A.Z., and V.K. conducted all experiments. D.K., S.K. and B.L.D. contributed to experimental measurements and data processing. K.K. and D.K. performed numerical simulations. A.Z. and D.Y.C. fabricated the samples. M.Z. contributed with the laser system for the experiment. G.Z., K.K., S.K., S.M. and Y.K. discussed results and wrote the manuscript.

# Supplementary Information

**S1. Polarization of high harmonics**

Figure S1 shows the dependence of the HHG intensity on the polarization angle of a linearly polarized pump for a certain orientation of the metasurface and for τ=100 fs.

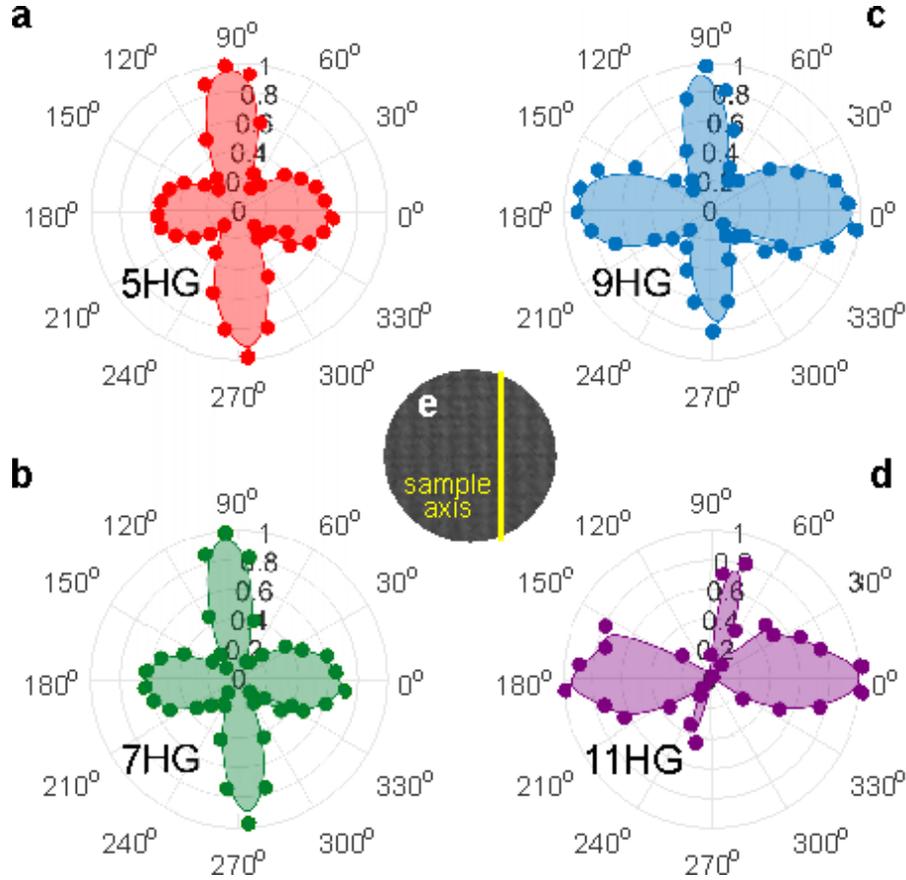

Fig. S1. Dependence of HHG intensity on the polarization angle of the linearly polarized pump at τ=100 fs for a certain orientation of the metasurface. The central image is obtained by SEM.

**S2. Temporal coupled-mode theory for BICs**

In this section we use the Gaussian convention for units. We derive the temporal coupled-mode theory equation, which describes the evolution of the resonance amplitude under arbitrary excitation. The total electric field $\mathbf{E}(\mathbf{r},t)$ can be separated into a sum of the background field $\mathbf{E}_{bg}(\mathbf{r},t)$ and the scattered field $\mathbf{E}_{sc}(\mathbf{r},t)$. The background field satisfies the Helmholtz's equation for the free space with substrate, but without the metasurface

$$\nabla^2 \mathbf{E}_{bg} - \frac{\varepsilon_{bg}(\mathbf{r})}{c^2} \frac{\partial^2}{\partial t^2} \mathbf{E}_{bg} = 0$$

The scattered field can be calculated via $\mathbf{E}_{bg}(\mathbf{r},t)$ and dyadic Green's function of the resonant system $\mathbf{G}_{EE}$

$$\mathbf{E}_{sc}(\mathbf{r}, t) = -\int_{-\infty}^{t} dt' \int d^3 r' [\epsilon(\mathbf{r}') - \epsilon_{bg}(\mathbf{r}')] \hat{\mathbf{G}}_{EE}(\mathbf{r},\mathbf{r}', t-t') \cdot \frac{i}{c} \frac{\partial}{\partial t'} \mathbf{E}_{bg}(\mathbf{r}', t')$$

For $\mathbf{r}$ and $\mathbf{r}'$ inside the metasurface, the electric-electric component of the dyadic Green's function can be expressed as a sum of contributions of resonant states $\mathbf{E}_j(\mathbf{r}, t)$ with complex frequencies $\omega_j - i\gamma_j$ [S1]

$$\hat{\mathbf{G}}_{EE}(\mathbf{r},\mathbf{r}', t-t') = -ic \sum_j \mathbf{E}_j(\mathbf{r}) \otimes \mathbf{E}_j(\mathbf{r}') e^{-i(\omega_j - i\gamma_j)(t-t')}$$

Here, the resonant states of the metasurface are normalized as follows [S1]

$$1 = \int_V dV \varepsilon(\mathbf{r}) \mathbf{E}_j \cdot \mathbf{E}_j - \int_V dV \mathbf{H}_j \cdot \mathbf{H}_j + i \oint_{S_V} (\mathbf{E}_j \times \mathbf{H}'_j - \mathbf{E}'_j \times \mathbf{H}_j) \cdot d\mathbf{S}$$

By inserting the Green's function into the equation for $\mathbf{E}_{sc}(\mathbf{r}, t)$, we obtain the expansion for the scattered field

$$\mathbf{E}_{sc}(\mathbf{r}, t) = -\sum_j \mathbf{E}_j(\mathbf{r}) e^{-i(\omega_j - i\gamma_j)t} \int_{-\infty}^{t} dt' e^{i(\omega_j - i\gamma_j)t'} \int d^3 r' [\epsilon(\mathbf{r}') - \epsilon_{bg}(\mathbf{r}')] \mathbf{E}_j(\mathbf{r}') \cdot \frac{\partial}{\partial t'} \mathbf{E}_{bg}(\mathbf{r}', t')$$

In general, $\mathbf{E}_{sc}(\mathbf{r},t)$ can be directly expanded into independent contributions of the metasurface resonances with resonant amplitudes $a_j$

$$\mathbf{E}_{sc}(\mathbf{r}, t) = \sum_j a_j(t) \mathbf{E}_j(\mathbf{r})$$

$$a_j(t) = e^{-i(\omega_j - i\gamma_j)t} \int_{-\infty}^{t} dt' e^{i(\omega_j - i\gamma_j)t'} s_j(t')$$

The excitation amplitudes $s_j(t)$ are defined as

$$s_j(t) = -\int d^3 r' [\epsilon(\mathbf{r}') - \epsilon_{bg}(\mathbf{r}')] \mathbf{E}_j(\mathbf{r}') \cdot \frac{\partial}{\partial t} \mathbf{E}_{bg}(\mathbf{r}', t)$$

Each of the resonant amplitudes satisfies the temporal coupled mode theory equation

$$\frac{d}{dt}a_j(t) = -i(\omega_j - i\gamma_j)a(t) + s_j(t)$$

We consider the excitation oscillating with the frequency ω close to the quasi-BIC frequency $\omega_0$, therefore the scattered field is dominated by one term with j=0. The total field inside the metasurface can be found as

$$\mathbf{E}(\mathbf{r},t) = \mathbf{E}_{bg}(\mathbf{r},t) + \mathbf{E}_0(\mathbf{r})e^{-i(\omega_0-i\gamma_0)t}\int_{-\infty}^{t}dt'e^{i(\omega_0-i\gamma_0)t'}s_0(t)$$

We consider the case with $\mathbf{E}_{bg}(\mathbf{r},t)$ being a pulse with a full width at half maximum of $\tau_{FWHM}$ ~ 100fs. We introduce the time parameter $\tau_{sp}=d/c$, where d is the characteristic spatial dimension of the metasurface, and the quasi-BIC lifetime $\tau_0=1/(2\gamma_0)$. For the experimental conditions, d ~ 1 μm, $\gamma_0$ ~ 5THz, which gives $\tau_{sp}$ ~ 3 fs, $\tau_0$ ~ 100 fs. The spatial integration in $s_0(t)$ goes over the volume of the metasurface. Therefore, under the integrand $|\mathbf{r}|/c \sim \tau_{sp} \ll \tau_{FWHM}, \tau_0$, so we can neglect spatial evolution of the pulse compared to its temporal evolution. This means, we can neglect mixing of t and **r** and separate the variables of the background field

$$\mathbf{E}_{bg}(\mathbf{r},t) = \mathbf{E}_{bg}(\mathbf{r})u(t)$$

Then, the excitation amplitude $s_0(t)$ is

$$s_0(t) = D_0 \dot{u}(t)$$

Here, $D_0$ is the amplitude of coupling between the background field and the quasi-BIC field. Due to reciprocity, $D_0$ is proportional to the square root of the rate of radiative losses of the quasi-BIC $\gamma_{0,rad}$

$$D_0 = (\gamma_{0,rad})^{1/2}\kappa_0$$

$$\kappa_0 = -(\gamma_{0,rad})^{-1/2}\int d^3r[\epsilon(\mathbf{r}) - \epsilon_{bg}(\mathbf{r})]\mathbf{E}_0(\mathbf{r})\cdot\mathbf{E}_{bg}(\mathbf{r})$$

The background field oscillates fast with the frequency $\omega \sim \omega_0 \gg 1/\tau_{FWHM}$, thus

$$\dot{u}(t) = -i\omega u(t)$$

Finally, we can write a compact expression for the resonant amplitude

$$a_0(t) = -i\omega D_0 \int_{-\infty}^{t}dt'u(t')e^{i(\omega_0-i\gamma_0)(t'-t)}$$

## S3. Resonant electric field for Gaussian pulse excitation

In this section we use the Gaussian convention for units. We consider excitation of the metasurface with a non-chirped Gaussian pulse, polarized along the x direction. The pulse is centered at $\omega \sim \omega_0$

$$u(t) = e^{-t^2/\tau^2} e^{-i\omega t}$$

$$\tau = \tau_{\text{FWHM}}/\sqrt{2\ln(2)}$$

The spatial dependence of the background field is given by

$$\mathbf{E}_{\text{bg}}(\mathbf{r}) = \mathbf{e}_x E_p^{(0)} \left[ \left( e^{-ik_0(z-z_{\text{sub}})} + re^{ik_0(z-z_{\text{sub}})} \right) H_\theta(z - z_{\text{sub}}) + te^{-ik_0 n_{\text{Si}}(z-z_{\text{sub}})} H_\theta(z_{\text{sub}} - z) \right],$$

where $z_{\text{sub}}$ is the position of the substrate, r and t is the reflection and transmission amplitudes of the substrate, and $H_\theta$ is the Heaviside theta function. By applying the temporal coupled-mode theory from section S2, we obtain the resonant amplitude

$$a_0(t) = -\frac{i\omega \tau \sqrt{\pi}}{2} D_0 e^{[\gamma_0 \tau - i(\omega - \omega_0)\tau]^2/4} e^{-i(\omega_0 - i\gamma_0)t} \operatorname{Erfc}\left[ \frac{\gamma_0 \tau - i(\omega - \omega_0)\tau}{2} - \frac{t}{\tau} \right].$$

The total field can be found as

$$\mathbf{E}(\mathbf{r},t) = \mathbf{E}_{\text{bg}}(\mathbf{r},t) + \mathbf{E}_0(\mathbf{r}) a_0(t).$$

## S4. Modeling of plasma generation

In this section and further we use the SI convention for units. Following Ref. [35] we calculate the electron density in the conduction band as:

$$\frac{\partial n}{\partial t} = R(t, \omega, |\mathbf{E}(t)|)$$

Here the strong field excitation rate R(t, λ, E) is calculated using the Keldysh expression

$$R(t, \omega, |\mathbf{E}(t)|) = 2 \frac{2\omega}{9\pi} \left( \frac{m^* \omega}{\hbar} \frac{\sqrt{1+\gamma^2}}{\gamma} \right)^{\frac{3}{2}} Q\left(\gamma, \frac{\Delta_{\text{eff}}}{\hbar \omega}\right) \times \exp\left[-\pi \left\langle \frac{\Delta_{\text{eff}}}{\hbar \omega} + 1 \right\rangle \frac{\mathcal{K}(\phi) - \mathcal{E}(\phi)}{\mathcal{E}(\Theta)} \right],$$

where

$$Q(\gamma, x) = \sqrt{\frac{\pi}{2\mathcal{K}(\Theta)}} \times \sum_{n=0}^{\infty} \exp\left[-\pi \cdot \frac{\mathcal{K}(\phi) - \mathcal{E}(\phi)}{\mathcal{E}(\Theta)} \cdot n\right] \cdot \Phi\left[\sqrt{\frac{\pi(\langle x+1 \rangle - x + n)}{2 \cdot \mathcal{K}(\Theta) \cdot \mathcal{E}(\Theta)}}\right],$$

K and E are complete elliptic integrals of first and second kind, respectively, with dimensionless derivatives

$$\Theta = \frac{1}{1+\gamma^2} \text{ and } \phi = \frac{\gamma^2}{1+\gamma^2},$$

$\gamma = \frac{\omega\sqrt{m^*\Delta}}{e|\mathbf{E}(t)|}$ is the Keldysh parameter, $\Delta$ is the intrinsic band gap, $m^*$ is the reduced mass of the effective electron and hole masses, $\omega$ is the laser frequency, $e$ is the elementary charge and $|\mathbf{E}(t)|$ is the amplitude of the total electric field inside the metasurface enhanced by the quasi-BIC resonance. Also,

$$\Phi(x) = \int_0^x \exp(\xi^2 - x^2) d\xi$$

is the Dawson integral, and

$$\Delta_{\text{eff}} = \frac{2}{\pi}\Delta \cdot \left[\frac{\sqrt{1+\gamma^2}}{\gamma}\mathcal{E}(\Theta)\right]$$

is the bandgap value modified due to the Stark shift. The parameters of silicon are $m^*=0.18m_e$ and direct bandgap $\Delta=3.4$ eV. For the simulations we use a pulse with the peak field

$$E_p^{(0)} = \left(\frac{2I_p}{c\varepsilon_0}\right)^{1/2}.$$

The peak intensity is $I_p = 3\times10^{11}$ W/cm$^2$ (pulse energy 0.8 µJ), which gives $E_p^{(0)} = 1.5\times10^9$ V/m. The value of the Keldysh parameter for $|\mathbf{E}(t)|=E_p^{(0)}$ is $\gamma = 0.6$. The pulse is resonant with the quasi-BIC wavelength 3.9 µm, $\tau_{\text{FWHM}} = 100$ fs, $\tau = 85$ fs. The losses of quasi-BICs are given by $\gamma_0 = 5$THz, which corresponds to the mode quality factor of 50. To find the Keldysh parameter, we evaluate $|\mathbf{E}(t)|$ at the electric hotspot of the quasi-BIC ($\mathbf{r} = \mathbf{r}_h$)

$$|\mathbf{E}(t)| = \left|E_{\text{bg},x}(\mathbf{r}_h)e^{-t^2/\tau^2} + E_{0,x}(\mathbf{r}_h)D_0 \cdot \left(-\frac{i\omega_0\tau\sqrt{\pi}}{2}\right)e^{\gamma_0^2\tau^2/4}e^{-\gamma_0 t}\operatorname{Erfc}\left[\frac{\gamma_0\tau}{2} - \frac{t}{\tau}\right]\right|$$

The numerical simulations are shown in Fig. S2. They demonstrate that the field strength of the laser pulse enhanced by the quasi-BIC reaches $3\times10^9$ V/m, which allows the density of carriers of $n_p = 6.8\times10^{19}$ cm$^{-3}$. Since we evaluated the field enhancement at the quasi-BIC electric hotspot, the value of $n_p=6.8\times10^{19}$ cm$^{-3}$ gives us the upper bound for the electron-hole plasma density. The lower bond in given by the density of carries induced solely by the incident laser pulse, which is $1\times10^{17}$ cm$^{-3}$

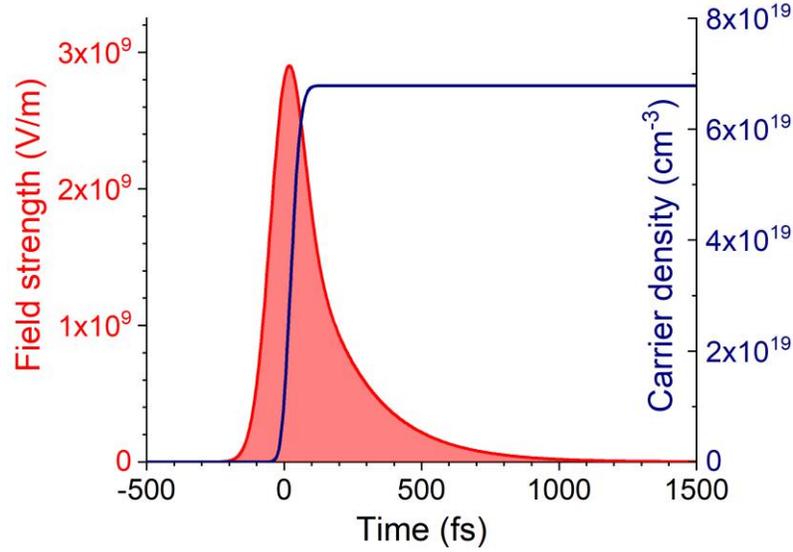

Fig.S2. Calculated resonant field strength $|E(t)|$ and electron-hole plasma carrier density for the 0.8 µJ pulse.

**S5. Calculations for electron-hole plasma contribution to dielectric permittivity of silicon**

According to the work [S2], electron-hole plasma changes silicon dielectric permittivity via Drude contribution, which can be described by the following formulas:

$$\Delta\varepsilon_D = \text{Re}[\Delta\varepsilon] + i\,\text{Im}[\Delta\varepsilon]$$

$$\text{Re}[\Delta\varepsilon] = -\frac{\omega_p^2}{\omega^2 + \tau_D^{-2}}, \quad \text{Im}[\Delta\varepsilon] = \frac{\omega_p^2 \tau_D}{\omega(1 + \omega^2 \tau_D^2)}$$

$$\omega_p^2 = \frac{n_p q^2}{\varepsilon_0 m^*}.$$

Here, $n_p$ is the plasma density, q is the elementary charge, $\varepsilon_0$ is the vacuum permittivity, $m^* = 0.18*m_e$ is the reduced mass of the effective electron and hole masses, $\tau_D$ is the Drude relaxation time. For low plasma densities, $\tau_D$ is determined by electron-phonon scattering process ($\tau_D=\tau_{ep}=100$ fs), while at high densities $\tau_D$ corresponds to electron-electron scattering process ($\tau_D=\tau_{ee}=1$ fs). The simulated change of refractive index and extinction coefficient vs. plasma density is shown in Figs. 3c-e of the main text.

## S6. Extinction coefficient of the amorphous silicon

In order to measure optical properties of the silicon films in our experiments, we used the ellipsometer RC2 by J.A. Woollam Ellipsometry Solutions. The result for imaginary part of refractive index (extinction coefficient) is shown in Fig. S3.

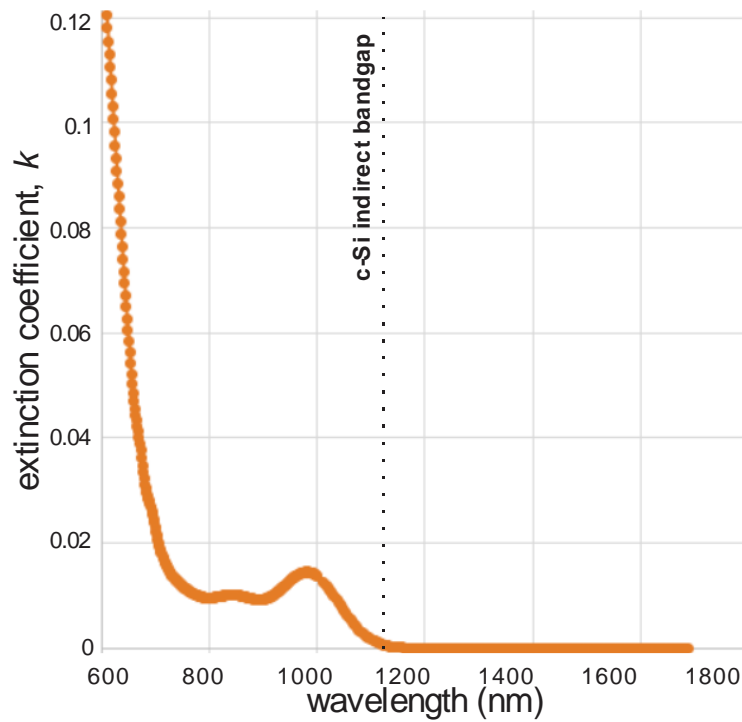

Fig. S3. Experimental data for the extinction coefficient of amorphous silicon.